\documentclass[journal, transmag]{IEEEtran}
\usepackage{cite}
\usepackage[pdftex]{graphicx}
\usepackage{subfigure}
\usepackage{amsmath}
\usepackage[caption=false,font=normalsize,labelfont=sf,textfont=sf]{subfig}
\usepackage[switch]{lineno}
\usepackage{multirow}
\begin{document}
\pagenumbering{gobble}
\title{General purpose readout board $\pi$LUP: overview and results.}

\author{{\textit{Nico~Giangiacomi}, Fabrizio~Alfonsi, Gabriele d'Amen, Gabriele~Balbi, Davide~Falchieri,}\\{Alessandro~Gabrielli, Giuseppe~Gebbia, Giuliano~Pellegrini, Davide~Soverini}
\thanks{ Manuscript received June 22, 2018  }
\thanks{N. Giangiacomi is with Istituto Nazionale di Fisica Nucleare (INFN) and Department of Physics and Astronomy (DIFA), Universit\'a di Bologna, Viale Berti Pichat 6/2, 40127 - Bologna - Italy. (e-mail: nico.giangiacomi@bo.infn.it)}
\thanks{F. Alfonsi, G. d'Amen, A. Gabrielli, G. Gebbia and D. Soverini are with INFN Bologna and University of Bologna.}
\thanks{G. Balbi, D. Falchieri and G. Pellegrini and  are with INFN Bologna.}
}

\maketitle

\begin{abstract}
This work gives an overview of the PCI-Express board $\pi$LUP, focusing on the motivation that led to its development, the technological choices adopted and its performance. The $\pi$LUP card was designed by INFN and University of Bologna as a readout interface candidate to be used after the Phase-II upgrade of the Pixel Detector of the ATLAS and CMS experiments at LHC. The same team in Bologna is also responsible for the design and commissioning of the ReadOut Driver (ROD) board - currently implemented in all the four layers of the ATLAS Pixel Detector (Insertable B-Layer, B-Layer, Layer-1 and Layer-2) - and acquired in the past years expertise on the ATLAS readout chain and the problematics arising in such experiments. 
Although the $\pi$LUP was designed to fulfill a specific task, it is highly versatile and might fit a wide variety of applications, some of which will be discussed in this work. Two 7$^{th}$-generation Xilinx FPGAs are mounted on the board: a Zynq-7 with an embedded dual core ARM Processor and a Kintex-7. The latter features sixteen 12.5$\,$Gbps transceivers, allowing the board to interface easily to any other electronic board, either electrically and/or optically, at the current bandwidth of the experiments for LHC. Many data-transmission protocols have been tested at different speeds; results will be discussed later in this work. 
Two batches of $\pi$LUP boards have been fabricated and tested; two boards in the first batch (version 1.0) and four boards in the second batch (version 1.1), encapsulating all the patches and improvements required by the first version.

\end{abstract}

\begin{IEEEkeywords}
ATLAS, FPGA, Detector Control System, Pixel Detector, Read-Out Driver, TDAQ.
\end{IEEEkeywords}

\IEEEpeerreviewmaketitle

\section{Introduction}
\label{sec:Introduction}

\IEEEPARstart{I}n the next few years, the Large Hadron Collider (LHC) at CERN will extend its investigation of the fundamental structure of physical matter by undergoing a series of major upgrades \cite{LHC}. All the experiments located on LHC circumference (such as the ATLAS experiment) will be upgraded as well, in order to reach the performance required for the physics challenges foreseen for the future. 
The key factors that will affect all the detectors are two: the increase of instantaneous luminosity - corresponding to an increase of the simultaneous collisions and hence of the amount of total data per time unit - and of the trigger rate, that will be on the order of 1$\,$MHz, ten times higher than the current rate of $\approx$100$\,$kHz. The combination of those two factors constitutes a major challenge for the electronic readout systems, since it directly effects the total throughput, i.e. the amount of data transmitted per time unit. Hence all the readout systems will have to provide a higher total bandwidth, capable of coping with the increased data throughput.

The purpose of this paper is to introduce a new readout card, called PIxel detector Luminosity UPgrade board ($\pi$LUP), developed by a joint effort from University and INFN of Bologna as a proposed readout upgrade system for the ATLAS experiment. The Bologna $\pi$LUP was designed as a natural upgrade of the current ATLAS Pixel Detector readout chain \cite{readout}, mainly composed of two electronic cards: Back Of Crate (BOC) \cite{BOC} - responsible for handling the control interface to the detector and the data from the detector - and the ReadOut Driver (ROD) \cite{ROD} - responsible of data processing and packaging. The ROD and BOC boards are connected together through a Versa Module Eurocard (VME) crate and together provide a total bandwidth of 5.12 Gbps. On the other hand, the $\pi$LUP board abandoned the VME connector, moving towards the solution of a 8 lanes Peripheral Component Interconnect Express (PCIe) bus. By exploiting the most recent technologies, it also merges in a single board both the I/O and the data processing capabilities and virtually provides a total bandwidth of 80 Gbps. 

\begin{figure}[htbp]
\centering
\includegraphics[width=\linewidth]{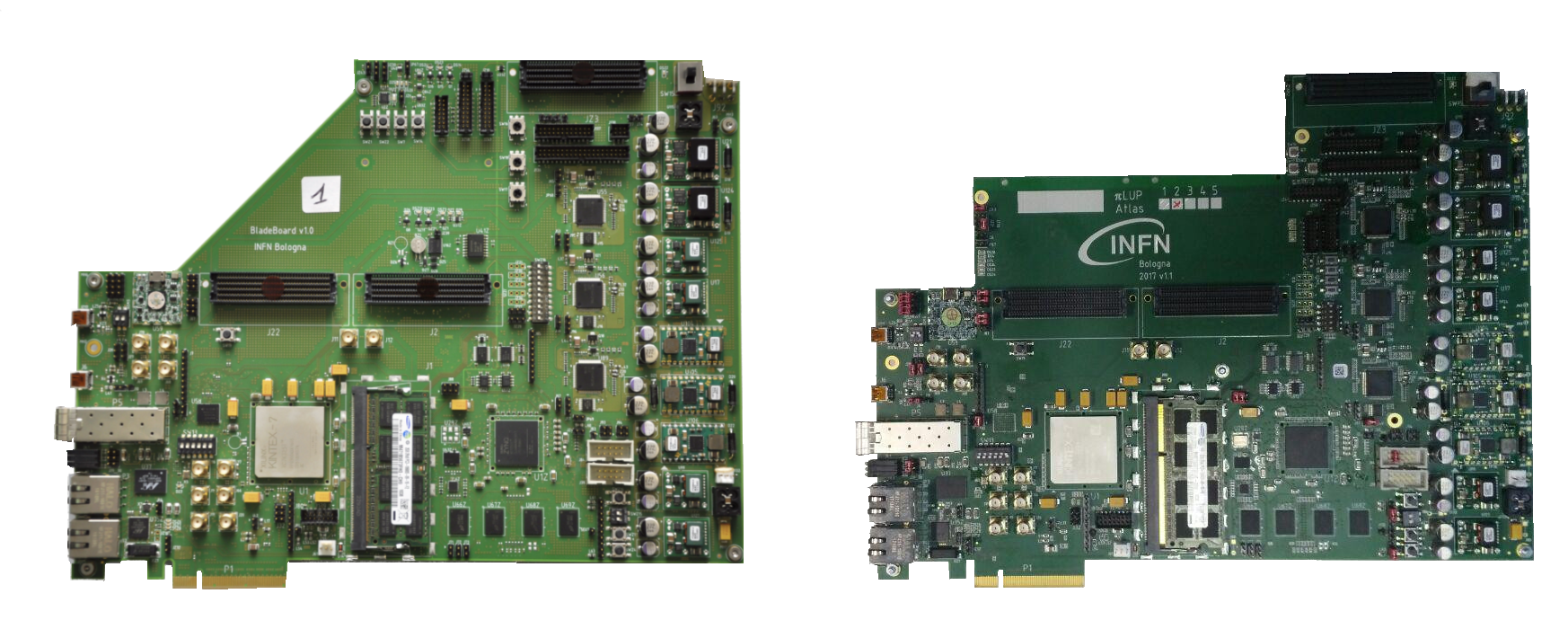}
\caption{Bologna $\pi$LUP v1.0 board (left) and $\pi$LUP v1.1 (right). The shape was modified to fit all the FMC Mezzanines.}
\label{fig:PiLUPs}
\end{figure}

Apart from the PCIe, the $\pi$LUP card features a huge variety of I/O connectors, such as two Universal Asynchronous Receiver-Transmitter (UART) ports, one 1$\,$Gbps Ethernet port, one 10$\,$Gbps Ethernet port, one Small Form-factor Pluggable (SFP+) connector and three FPGA Mezzanine Card (FMC) connectors. Having a wide choice of different I/O interfaces attributes a great versatility to the $\pi$LUP card, making it perfectly suited to act as a general purpose readout board. In fact, although it was designed to fulfill a specific task, it can be used to interface several types of front-end chips or electronic systems.

Two first prototypes of Bologna $\pi$LUPs (version 1.0) were produced in 2016. Most of the I/O connectors and the internal functionalities were successfully tested. However, some small patches were required and the shape of the board had to be revisited to properly fit one of the FMC connectors. 
Those revisions led to the fabrication of four new boards (version 1.1) in 2018; the two version are shown in Figure \ref{fig:PiLUPs}.

In the following sections a technical overview on the main components of the $\pi$LUP board will be presented, as well as its possible applications in different projects and the results obtained.

\section{$\pi$LUP board overview}

\begin{figure}[htbp]
\centering
\includegraphics[width=\linewidth]{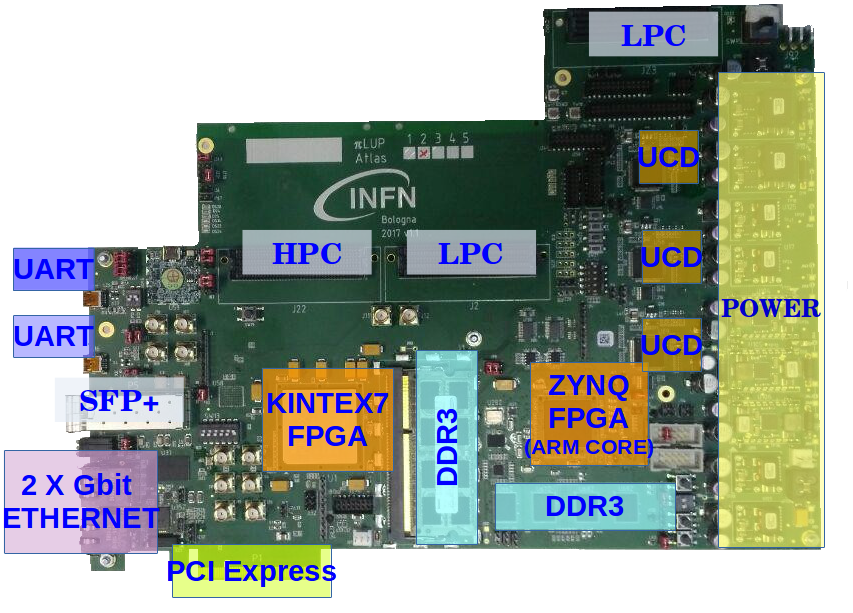}
\caption{Bologna $\pi$LUP v1.1 board. FPGAs, I/O ports and other main components are highlighted in the picture.}
\label{fig:Pilup_labels}
\end{figure}

The Bologna $\pi$LUP card is a 16 layers PCI Express board capable of interfacing to several different other boards or front-ends and processing data at high speed. Figure \ref{fig:Pilup_labels} shows the main components on the board. 

The $\pi$LUP features two Xilinx 7$^{th}$ series FPGAs arranged in a Master/Slave architecture and connected together by a bus - namely KZbus - composed of 5 single ended and 21 differential lines. A Zynq XC7Z020-1CLG484C - embedding a physical dual-core ARM Cortex-A9 processor - is the Master FPGA and is in charge of controlling the data-flow and status of the Slave FPGA, a Kintex XC7K325T-2FFG900C.
The Kintex device handles all the high speed I/O communications through 16 internal physical transceivers (GTx) \cite{Transceivers} running at up to 12 Gb/s. The transceivers are connected to different types of physical ports such as the 8 lanes PCI Express, FMCs, SFP+ and 10$\,$Gbit-Ethernet as described in Table \ref{tab:gtx}. There are three FMC connectors on the board; one of them, a Low Pin Count (LPC) FMC connector is connected to the Zynq FPGA and the others, a High Pin Count (HPC) FMC connector and another LPC FMC connector are connected to the Kintex 7 FPGA. Those connectors are built accordingly to the VITA Standard 57.1 and can support any FMC mezzanine that respects the same standard. 

\begin{table}[htbp]
 \centering
 \normalsize 
 
\caption{GTx transceiver and reference clocks connections in $\pi$LUP Kintex-7 FPGA }
\begin{tabular}{ |c|c|c|c|c| } 
\hline
Bank & REFCLK0 & REFCLK1 & MGT & I/O Port \\
\hline
\hline
\multirow{4}{2em}{115} & \multirow{4}{4em}{} & \multirow{4}{4em}{\centering PCIe REFCLK}  & 0 & PCIe lane 7 \\
& & &  1 & PCIe lane 6 \\
& & &  2 & PCIe lane 5 \\ 
& & &  3 & PCIe lane 4 \\
\hline
\multirow{4}{2em}{116}  & \multirow{4}{4em}{\centering SMA REFCLK} & \multirow{4}{4em}{\centering LPC REFCLK}   & 0 & PCIe lane 3 \\
& & &  1 & PCIe lane 2 \\
& & &  2 & PCIe lane 1 \\
& & &  3 & PCIe lane 0 \\
\hline
\multirow{4}{2em}{117}  & \multirow{4}{4em}{\centering 125 MHz ck source} & \multirow{4}{4em}{\centering Si5326}   & 0 & SMA \\
& & &  1 & Gb-Ethernet \\
& & &  2 & SFP+ \\
& & &  3 & FMC LPC \\
\hline
\multirow{4}{2em}{118}  & \multirow{4}{4em}{\centering HPC REFCLK 0} & \multirow{4}{4em}{\centering HPC REFCLK 1}   & 0 &  FMC HPC 0 \\ 
& & &  1 & FMC HPC 1 \\
& & &  2 & FMC HPC 2 \\
& & &  3 & FMC HPC 3 \\
\hline
\end{tabular}
\label{tab:gtx}
\end{table}

\subsection{Clock Distribution}

Several clock sources are present in the board. The Zynq-7 FPGA is associated to three main clock sources. A 200 MHz system clock is provided by a SiTime SiT9102, a differential output programmable oscillator providing ±10 ppm frequency stability with sub-piscosecond phase jitter; a configurable user clock is provided by a Silicon Labs Si570, a low jitter oscillator that supports frequencies between 10 and 1400$\,$MHz; the Processing System (PS) clock is provided by a 50$\,$ppm 33.33$\,$MHz oscillator. The Kintex-7 device features another 200 MHz SiT902 system clock and programmable Si570 user clock, as well as other clock inputs required by the GTx transceivers as reference clocks \cite{Transceivers}. Some of those reference clocks are embedded on the $\pi$LUP itself, while others must be provided from the outside.
The two sources provided by the $\pi$LUP are a 125$\,$MHz Ethernet reference clock, provided by the combination of a 25$\,$MHz crystal oscillator and a Integrated Device Technology 844021I-01 crystal oscillator interface, and a programmable reference clock, provided by a Silicon Labs Si5326 jitter cleaner. The external reference clock sources must be provided by the PCIe connector, by the LPC and HPC FMC connectors or by the SMA connectors. Table \ref{tab:gtx} shows how the reference clocks are associated to the different GTx transceivers.

\section{Software architecture}
\begin{figure}[htbp]
\centering
\includegraphics[width=\linewidth]{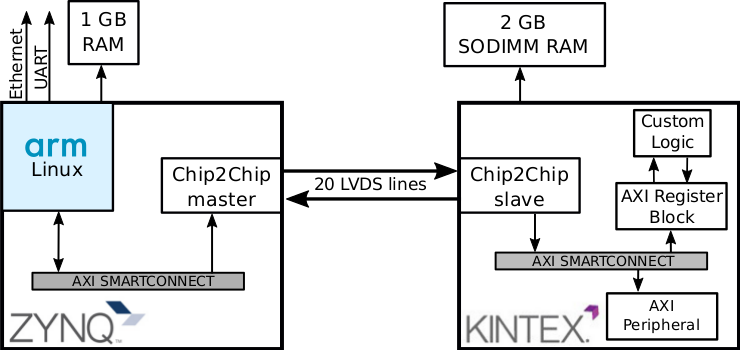}
\caption{Diagram of the $\pi$LUP control infrastructure.}
\label{fig:Software_architecture}
\end{figure}

The two FPGAs present on the board are intended to be used in a master-
slave configuration, with the Zynq, or more precisely its ARM-based Processing
System (PS), controlling any peripheral in the system and acting as a main
interface to the user. A diagram of the setup is shown in Figure \ref{fig:Software_architecture}.
Inside the Zynq, the PS communicate with the FPGA through the AMBA AXI
protocol. This channel is extended to the Kintex by the AXI Chip2Chip IP
core offered by Xilinx. This core transparently bridges a 32-bit AXI bus to the
slave device so that any peripheral present in the Kintex can be addressed from
the ARM as if it was directly implemented in the Zynq. The physical interface
is quite flexible and can be adapted to a limited pin count; in this case the
communication employs 20 differential lines operating at 200 MHz Double Data Rate DDR (9 data
bits plus clock for each direction). On startup the Chip2Chip automatically
performs a deskewing self-calibration and then is immediately ready to use.
In any configuration the C2C master shows a single AXI slave port and the
C2C slave a single AXI master port, so the bridge is not exactly “symmetrical”,
but this do not entail a limitation in this design. Four interrupt ports for each
direction are also present. The C2C channel multiplexer assigns higher priority
to those over AXI data.
The Zynq PS runs an embedded Linux distribution generated with the Xilinx Petalinux
tools, providing a high level interface to any functionality present in the board
(including web services such as an SSH server). During the boot-up, the
Linux image can be loaded from the on-board flash chip or downloaded from a
remote server with the Trivial File Transfer Protocol (TFTP) protocol.
Generally most AXI cores offered by Xilinx also ship a driver often included
in the Linux kernel tree. For custom-made cores without an AXI interface, a
control interface is offered by an AXI-addressable register block, that is directly
accessed from Linux user-space using the generic-User Input Output (UIO) driver. The UIO driver
greatly simplifies the development of drivers that does not require a custom kernel
module and fits very well with the view of offering a higher level interface to
the functionalities implemented in the FPGA.
Others off-chip devices, such as the I2C-programmable Si570 clock generator,
Si5326 PLL and bus multiplexer, can also be controlled directly from Linux by means of an AXI-based
I2C controller. The kernel already includes drivers for the bus multiplexer and
the Si570; the former transparently manages the multiplexer and the kernel is
simply presented with a number of buses than can be directly accessed. In this
application the Si5326 is programmed by a custom user-space software that
calculates the required values of its internal registers and write them with a
simple file access to the character devices representing the muxed bus associated
to the device.

\section{Applications for the $\pi$LUP board}

As already stated in Sec. \ref{sec:Introduction}, the $\pi$LUP board was designed to fulfill a specific task, i.e. the readout upgrade for the next generation ATLAS Pixel Detector, merging in a single board both I/O connections and data processing. Nevertheless, the $\pi$LUP features a huge variety of I/O connectors and three FMC connectors, making the board highly versatile and able to interface a wide variety of different other electronic devices and Front-End chips. The choice of having two FPGAs connected in a Master/Slave mode guarantee enough power to perform high level control operation on the board (Zynq-7 ARM core) and handle I/O communications through several different protocols (Kintex 7) while at the same time maintaining a relatively low price.

The main possible applications for the $\pi$LUP board are three:
\begin{itemize}
  \item readout control system; the $\pi$LUP can be used to directly interface a front-end device performing data-processing, data transfer to the PC via PCIe bus, online on-chip histogramming and system control. The maximum bandwidth in this scenario is limited by the PCIe data transfer rate, i.e. 4 GB/s for the 8 lane gen.$\,$2 PCIe bus or 7.9 GB/s for a 8 lane gen.$\,$3 PCIe bus;
  
  \item data generator/ front-end emulator; the $\pi$LUP can be used to generate/emulate data to be sent to other systems, for example to validate a reconstruction or data-processing algorithm. The maximum bandwidth in this case is 10 GB/s, which is the maximum speed of the 8 GTx transceivers not used in the PCIe bus;
  
  \item bridge between two different systems; the $\pi$LUP can be used as a bridge to connect two different readout systems that use different protocols or different communication physical layers. The maximum bandwidth in this scenario is highly influenced by the interfaced systems.
\end{itemize}

\subsection{Interface with Felix}

\begin{figure}[htbp]
\centering
\includegraphics[width=\linewidth]{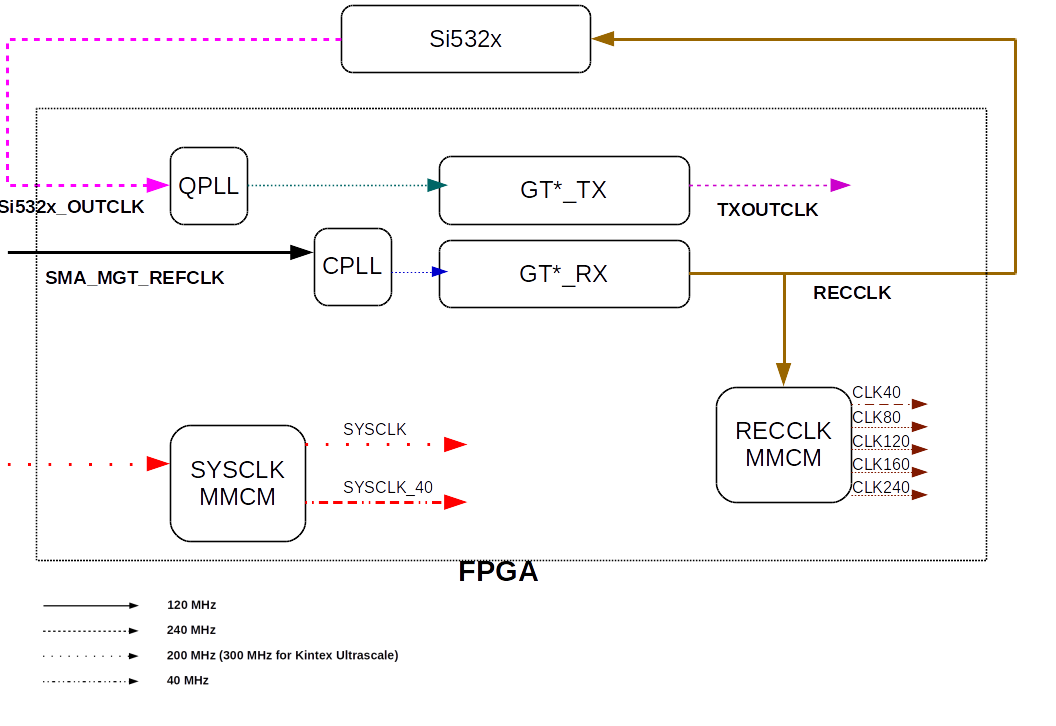}
\caption{Clock distribution of the $\pi$LUP GBT/Full Mode protocol firmware. }
\label{fig:Clock_scheme}
\end{figure}

A first proof of the many possibilities of the $\pi$LUP board came from a integration test with Felix boards from the Felix Project \cite{Felix}.
The $\pi$LUP was connected through optical fiber to a Mini-Felix card (Xilinx VC709 evaluation board \cite{VC709}) and a FLX-712 card \cite{Felix}. The test showed that the two boards were able to establish a communication via both GigaBit Transceiver (GBT, 4.8 Gb/s) \cite{GBT} and custom Felix Full Mode (9.6 Gb/s) protocols. 
For both configurations, the $\pi$LUP used the Common Phase Locked Loop (CPLL) of the transceivers to recover the clock from the incoming data stream; the clock was then cleaned using the jitter cleaner Si5326 on the board and propagated to the Quad PLL (QPLL) for the GTx transmitters, creating a
synchronous data acquisition system. Figure \ref{fig:Clock_scheme} shows the clock distribution of the $\pi$LUP board.

Using a Faster Technology FM-S14 FMC HPC Mezzanine Card (shown in Figure \ref{fig:FM-S14}), providing four additional SFP+ connectors, four link connections were simultaneously established between the $\pi$LUP and the Felix cards, resulting in a total throughput of 19.2 Gbps in GBT configuration and 38.4 Gbps in Full Mode configuration. Both configurations were tested for about one hour and no errors were found, demonstrating the reliability of the connections.

\subsection{Interface with the RD53A front-end chip}

This section shows an example of a real application for the $\pi$LUP board, used in collaboration with the Felix Project to interface the Felix card to the new generation pixel front-end chip: RD53A \cite{RD53A}. 

The need of use the $\pi$LUP board as an interface system arise from the physical and protocol incompatibilities between the Felix card and the RD53A chip. The first communicates via optical fibers through either 4.6 Gbps GBT or 9.8 Gbps Full Mode protocols, while the latter - currently bonded on a custom designed PCB called Single Chip Card (SCC) - communicates via Display Port (DP) connectors through 160 Mbps E-link (input) and four lanes 1.28 Gbps Aurora 64/66 protocol (output).  

\begin{figure}[htbp]
\centering
\includegraphics[width=\linewidth]{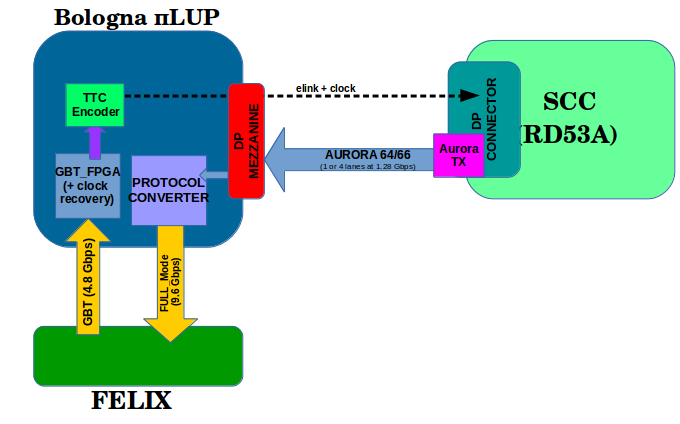}
\caption{Block Diagram showing the $\pi$LUP board as an interface between Felix and RD53A front-end chip..}
\label{fig:Protocol_converter}
\end{figure}

The role of the $\pi$LUP is hence to act as a bridge between this two systems, handling both the Felix-to-RD53A data-path (downlink) and the RD53A-to-Felix path (uplink). This is done through different firmware blocks, as shown in Figure \ref{fig:Protocol_converter}. The GBT$\_$FPGA block decodes the GBT-formatted data from Felix containing the configurations commands for the RD53A chip and also synchronizes to the Felix clock, recovering it from the data-stream. Both configuration commands and clock are then propagated to the TTC Encoder firmware block, which is in charge of converting the commands to a RD53A compatible format and of encapsulating them in a single 160 Mbps serial line, connected to one of the DP connector data lanes.

Concurrently the $\pi$LUP receives and decodes Aurora 64/66 data from the RD53A chip, coming from the other four data lanes of the DP connector. Those four lanes 1.28 Gbps data (resulting in a total throughput of 5.12 Gbps) are then passed to the Protocol Converter firmware block, which merges them in a single Full Mode stream that is transmitted to Felix via optical connection.  

\begin{figure}[htbp]
\centering
\includegraphics[width=\linewidth]{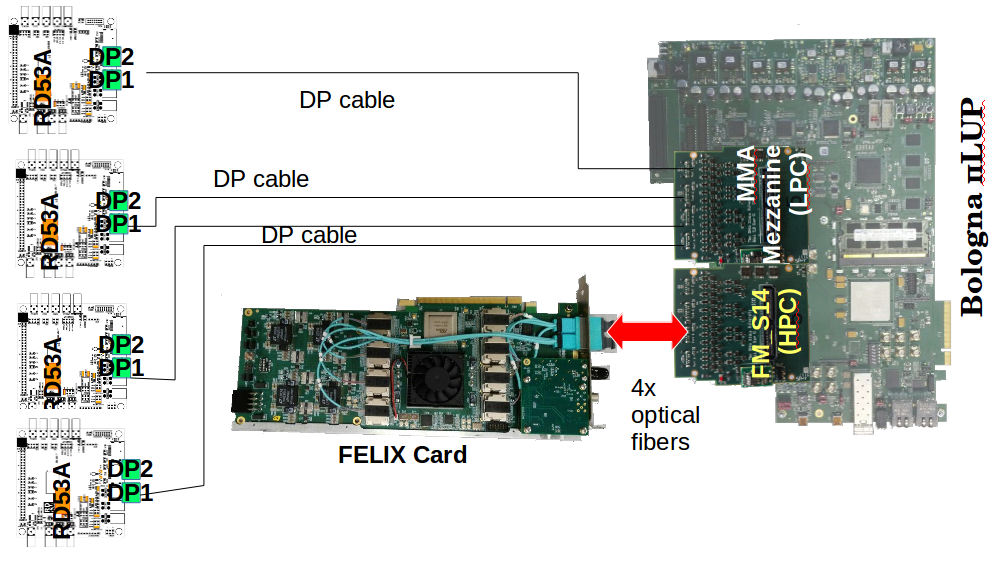}
\caption{Configuration between Felix and four RD53A modules, using the Bologna $\pi$LUP as an interface. This configuration requires the usage of the MMA FMC LPC mezzanine (four DP connectors) and the FM-S14 FMC HPC mezzanine (four SFP+ connectors).}
\label{fig:4_SCC}
\end{figure}

Although the $\pi$LUP doesn't include a DP connector in its design, the usage of FMC cards can sort through this problem. In particular, two custom FMC mezzanines were developed to be used as an interface to the RD53A: Single Module Adapter (SMA), a HPC FMC mezzanine featuring two DP connectors, and Multiple Module Adapter (MMA), a LPC FMC mezzanine featuring four mini-DP connectors.  

\begin{figure}[htbp]
\centering
\includegraphics[width=3cm]{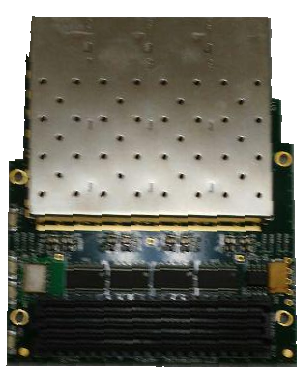}
\caption{FM-S14 FMC HPC card featuring two IDT ICS8N4Q001 programmable reference clocks and fours SFP+ connectors.}
\label{fig:FM-S14}
\end{figure}

The maximum throughput for the $\pi$LUP can be obtained by the usage of both the MMA LPC mezzanine and the FM-S14 HPC mezzanine (featuring four SFP+ connectors), shown in Figure \ref{fig:FM-S14}. Using this configuration, shown in Figure \ref{fig:4_SCC}, a Felix can interface four RD53A chips, resulting in a total throughput of $4\times5.12\,Gbps=20.48\,Gbps$. 

\section{Test results}
To evaluate and monitor the performance of the GTx transceivers on the $\pi$LUP board, the LogiCORE$^{TM}$ IP Integrated Bit Error Ratio Test (IBERT) core for 7 series FPGA \cite{IBERT} was used. This IPcore generates the eye diagrams and calculates the Open Area and Bit Error Rate (BER) for the different I/O interfaces, that were connected in loopback mode. To test the four transceivers in the FMC HPC connector, a Faster Technology FM-S14 Mezzanine Card was used.

This Mezzanine Card, shown in Fig. \ref{fig:FM-S14} implements four SFP+ connectors and two IDT ICS8N4Q001 programmable reference clocks.

\begin{figure}[htbp]
\centering
\includegraphics[width=\linewidth]{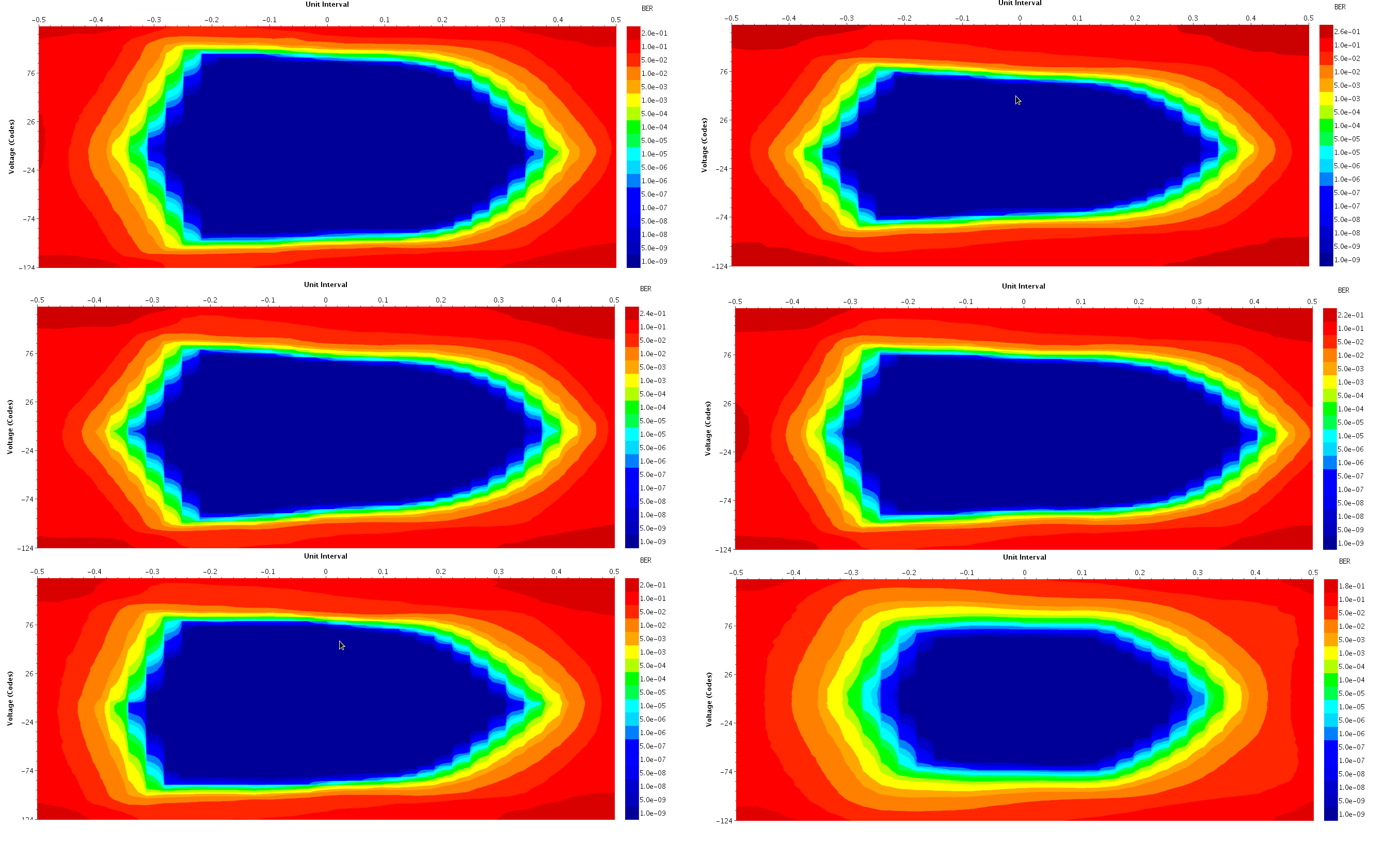}
\caption{Eye Diagram scan results run at 5 Gbps, PRBS 31bit and BER $10^{-9}$ of: FMT HPC MGT 0 (top left), FMT HPC MGT 1 (top right), FMT HPC MGT 2 (middle left), FMT HPC MGT 3 (middle right), SFP+ (bottom left) and SMA MGT (bottom right).}
\label{fig:OpenArea_5G}
\end{figure}

\begin{figure}[htbp]
\centering
\includegraphics[width=\linewidth]{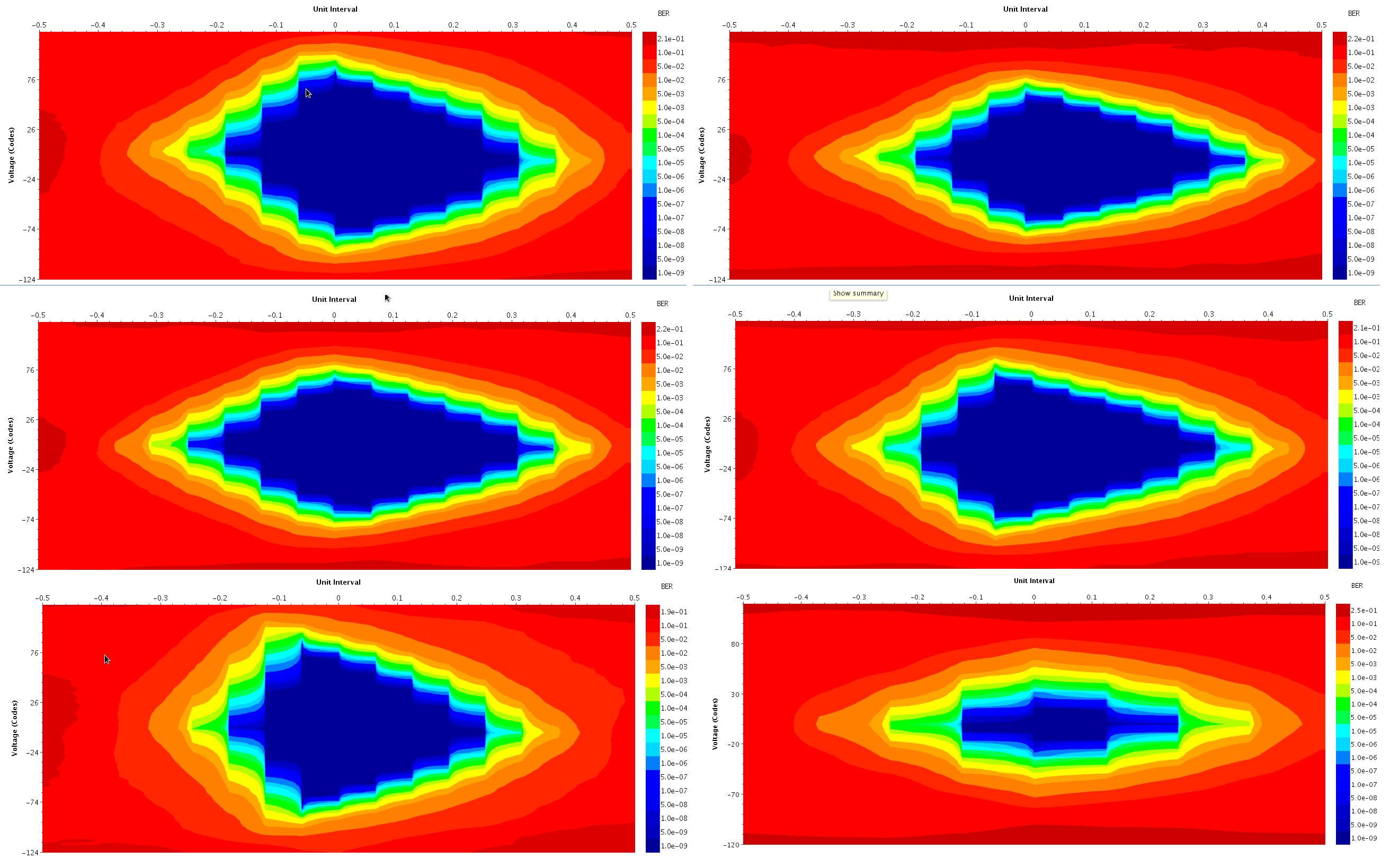}
\caption{Eye Diagram scan results run at 10 Gbps, PRBS 31bit and BER $10^{-9}$ of: FMT HPC MGT 0 (top left), FMT HPC MGT 1 (top right), FMT HPC MGT 2 (middle left), FMT HPC MGT 3 (middle right), SFP+ (bottom left) and SMA MGT (bottom right).}
\label{fig:OpenArea_10G}
\end{figure}

The BER and eye diagram scans were performed at 5$\,$Gbps and at 10$\,$Gbps; the two speeds were chosen to be slightly higher than the design operation mode protocol speeds, i.e. GBT (4.8$\,$Gbps) and Full Mode (9.6$\,$Gbps). The tests were performed using a PseudoRandom Binary Sequence (PRBS) of 31 bits and requiring a BER $\leq 10^{-9}$. The BER test was then continued until the error rate reached was $\leq 10^{-14}$. Figure \ref{fig:OpenArea_5G} shows the eye diagram of the tests run at 5$\,$Gbps and Fig. \ref{fig:OpenArea_10G} shows the results at 10$\,$Gbps; Table \ref{tab:openArea} shows the open area results.

\begin{table}[htbp]
 \centering
 \normalsize 
 
\caption{Open Area results run at 5/10$\,$Gbps, PRBS 31-bit and BERR $10^{-9}$}
\begin{tabular}{ |c|c|c| } 
\hline
I/O Connector &   Open Area  &   Open Area \\
               &  (5$\,$Gbps) &  (10$\,$Gbps) \\
\hline
\hline
FMT HPC MGT 0 & 11952 & 2784 \\
\hline
FMT HPC MGT 1 & 9008 & 2272 \\
\hline
FMT HPC MGT 2 & 10480 & 2512\\
\hline
FMT HPC MGT 3 & 11280 & 2480\\
\hline
SFP+ & 10432 & 2320 \\
\hline
SMA MGT & 6704 & 512\\
\hline
\end{tabular}
\label{tab:openArea}
\end{table}

\subsection{PCI Express}

\begin{figure}[htbp]
\centering
\includegraphics[width=\linewidth]{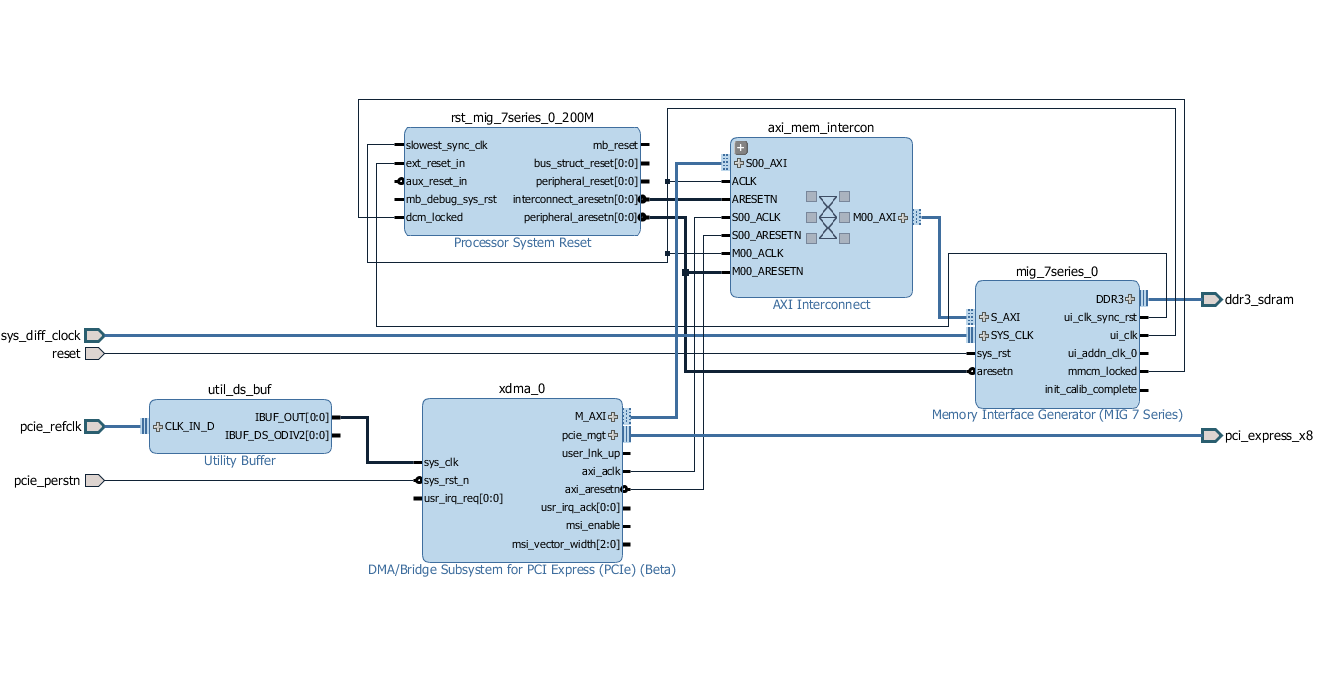}
\caption{Block Diagram of the PCI Express validation firmware, realized with Vivado$^{TM}$ IP Integrator. The main blocks are the DMA/Bridge subsystem for PCIe (XDMA) and the Memory Interface Generator (MIG).}
\label{fig:PCI_firmware}
\end{figure}

The workflow to validate and measure the performance of the PCIe Gen 2 bus on the $\pi$LUP required the design of a custom firmware implemented on the Kintex 7 FPGA and the development of custom Linux drivers allowing read and write operations from and to the RAM memory on the board, plugged in one of  the PCIe slots of Linux pc. The test design consisted in using the PCIe bus to write and read the 2 GByte DDR3 RAM associated to the Kintex device, measuring BER and speed. The firmware was entirely designed using the Vivado$^{TM}$ IP Integrator as it is shown Figure \ref{fig:PCI_firmware}. It is composed of a DMA/Bridge subsystem for PCIe (XDMA), a Memory Interface Generator (MIG) and other support logic needed to correctly connect these two blocks.

The XDMA is an IP block that implements a high performance, configurable Scatter Gather DMA for use with the PCIe Gen2.1 and Gen3.x that can be configured to be a bridge between the PCI Express and AXI memory spaces. The master side of this block reads and writes requests on the PCIe and its core enables the user to perform direct memory transfers, both Host to Card (H2C), and Card to Host (C2H).
The MIG IP core is a controller and physical layer for interfacing 7-series FPGA to DDR3 memory.

The custom drivers required to perform the test were developed for a Linux Ubuntu 16.04 Operative System. The test showed a peak user payload of 3.5 GBps data transfer when using buffers of 2 Mbyte and a BER$\leq10^{-14}$ corresponding to 24 TByte data transferred without errors. 
\section{Conclusions}

This paper presented the readout board $\pi$LUP designed in Bologna, focusing on its component and the result achieved. The technological choices and solutions were optimal to overcome many of the challenges presented by the upgrade of LHC. Hence, the $\pi$LUP is well suited to be used as an upgrade card for the readout system of one of the main LHC experiments, while at the same time maintaining a high flexibility and the potentialities to be used for several other applications, some of which were discussed in the previous sections. The outstanding results achieved and the relatively low cost make the board an interesting candidate for a high speed readout system.  


%





\ifCLASSOPTIONcaptionsoff
  \newpage
\fi

\end{document}